# Asymmetric Two-Terminal Graphene Detector for Broadband Radiofrequency Heterodyne- and Self-Mixing

*Jiayue Tong,[†] Matthew C. Conte,[‡] Thomas Goldstein,[†] Sigfrid K. Yngvesson,[‡]*

*Joseph C. Bardin[‡] and Jun Yan[†*]*

[†]Department of Physics, University of Massachusetts, Amherst, Massachusetts 01003, USA

[‡]Department of Electrical and Computer Engineering, University of Massachusetts, Amherst, Massachusetts 01003, USA

[*]Corresponding Author: Jun Yan.    Tel: (413)545-0853    Fax: (413)545-1691    E-mail: yan@physics.umass.edu

**Abstract**

Graphene, a single atomic layer of covalently bonded carbon atoms, has been investigated intensively for optoelectronics and represents a promising candidate for high-speed electronics. Here we present a microwave mixer constructed as an asymmetrically-contacted two-terminal graphene device based on the thermoelectric effect. We report a 50 GHz (minimum) mixer bandwidth as well as 130 V/W (163 mA/W) extrinsic direct-detection responsivity. Anomalous second-harmonic generation due to self-mixing in our graphene detector is also observed. Careful investigation of the responsivity from four different approaches gives consistent results, confirming the exceptional performance of our zero-bias device operating at room temperature. The 50 GHz bandwidth indicates an extremely fast response time and our experimental results represent an encouraging advance towards practical graphene microwave devices, with anticipated future applications extended through millimeter wave and terahertz frequencies.

**Manuscript text**

The interest in graphene has been rapidly growing in recent years with numerous demonstrations showing its usefulness as a practical industrial material. The roll-to-roll mass production of this layer-by-layer transferrable two-dimensional atomically thin sheet,[1] together with its unique properties,[2] make graphene highly attractive for integration with various waveguide technologies, silicon compatible integrated circuits, as well as large area flexible electronics.[3–7] A highly promising area of application for graphene is high-frequency large-bandwidth electronics based on characteristics such as its high intrinsic carrier mobility, high carrier saturation velocity, ultra-short response time and ambipolar charge transport.[8–10]

Graphene radiofrequency (RF) mixers and harmonic multipliers have been explored for about a decade.[5,10–16] Initial graphene mixers demonstrated the feasibility of designing such devices,[10] but had performance data that were orders-of-magnitude away from values required in applications.[9] The graphene field effect transistor ("GFET") configuration dominated the gradual development of graphene mixers through the GHz to hundreds of GHz frequency range.[5,10–16] At low frequencies, GFET mixers and multipliers still cannot compete with the dominant CMOS and other semiconductor technologies. However, at millimeter wave frequencies recent GFET mixers with conversion loss of 29 dB and bandwidth of 15 GHz[14] have begun to demonstrate comparable performances. This type of IC could be useful in the future development of high speed communication systems around 200 GHz.[17] At even higher frequencies, the GFET mixers have higher conversion loss of 60 dB at an LO power of -10 dBm, and smaller bandwidth of 5 GHz,[16] suggesting that the FET technology has problems due to parasitic reactance at these higher frequencies. Note that the GFET microwave and millimeter-wave mixers studied so far have consistently smaller bandwidth than what the intrinsic graphene speed has to offer. [5,12,14,16] In contrast, applications in high-speed (i.e. wide-bandwidth) data communications utilizing fiber and waveguide based optical systems, which take advantage of the ability of graphene to absorb and detect a very wide range of infrared photons at speeds higher than 60 GHz, have advanced more rapidly.[6,7,18–20]

In this Letter, we use a two-terminal asymmetrically-contacted graphene device to realize a high-performance RF heterodyne mixer with low conversion loss and very broad bandwidth, exceeding 50 GHz, which is the upper frequency limit of our present measurement system. Such a wide bandwidth is unprecedented for graphene RF mixers and represents an encouraging development upon several previous mixer designs whose best bandwidth was limited to 15 GHz. [5,12,14,16] Further, since the thermoelectric (TE) effect harnessed in the device is essentially frequency independent, we can extrapolate the performance of the TE mixer to millimeter wave and THz frequencies, which, combined with the highly-desirable broad bandwidth, may find useful future applications in emerging areas of high speed data communication and (sub)millimeter wave imaging.[17,21] We further found that our graphene device is highly efficient in generating second-harmonic signals. This is surprising because as a centrosymmetric material, graphene's second-order nonlinear susceptibility $\chi^{(2)}$ is expected to be zero.[22] We interpret our second-harmonic signal as being the result of RF self-mixing, due to the asymmetric contacts of our device. This interpretation is confirmed by a control sample with one order of magnitude smaller asymmetry, which correspondingly produces a second-harmonic signal that is approximately two orders of magnitude weaker.

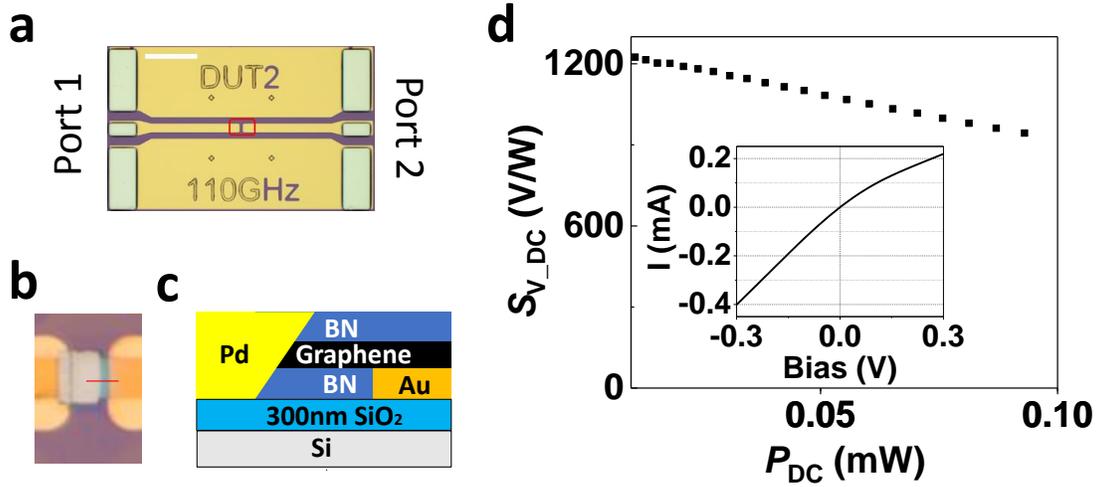

**Figure 1**. (a) Picture of the coplanar waveguide (CPW). The white scale bar is 100 μm. The red rectangle is the area where graphene is transferred to. (b) Zoomed-in view of the BN/graphene/BN piece transferred to the CPW. (c) Schematic sideview of the graphene device indicated by the red line in (b). (d) Responsivity calculated from the I-V curve. Inset: I-V curve of the device.

Our RF mixer device is made by dry-transfer of hexagonal boron nitride (hBN) protected graphene to the signal line of a pre-fabricated GSG (ground-signal-ground) coplanar waveguide (CPW; Fig. 1a). To break the mirror symmetry of the device, one of the two terminals is etched with $CHF_3$ and $O_2$ plasma, and subsequently contacted from the edge by palladium (Pd) sputtering (more fabrication details in Methods).[23] Figure 1b shows the optical microscope image of the sample, and the schematic side view is illustrated in Fig. 1c. Comparing with several other different contact strategies we explored before,[24] such as Au edge source / Pd top drain, Au edge source / Au top drain and Au top source / Pd top drain, we found that this combination gives lower impedance while maintaining high responsivity.

We first characterize the asymmetry of our device by performing DC charge transport measurements. As shown in the inset of Fig. 1d, the *I-V* curve of our device is nonlinear and asymmetric under positive and negative biases. This asymmetry arises from the thermoelectric response of the graphene device under DC current heating, which creates a hot spot in the center of the graphene sample and a temperature gradient towards the metal-contact heat sinks. The non-uniform temperature profile along the device and the asymmetric Fermi energy distribution due to the different contacts create a non-zero thermoelectric voltage superimposed on the applied bias voltage.[24,25] We make use of the *I-V* curve to extract the TE voltage responsivity of the device. When the sample is biased with a positive voltage $V$, the current $I_+$ in the sample is given by $V/R_G + I_{th}$, where $R_G$ is the resistance of the sample (about 800 Ω for this device) and $I_{th}$ is the thermoelectric current. When the bias voltage changes sign, so does the bias current, while $I_{th}$ remains in the same direction which is pre-determined by the sample asymmetry produced during device fabrication, i.e. $I_- = -V/R_G + I_{th}$. The thermoelectric voltage responsivity of the device can then be calculated as

$$S_{V\_DC} = I_{th} R_G / P_{DC} \qquad (1),$$

where $P_{DC} = V^2/R_G$ is the heating power. Figure 1d main panel shows the responsivity of the device as a function of $P_{DC}$, which is about 1000 ± 200 V/W; a similar magnitude of responsivity has been observed before in other graphene TE devices.[24,25]

The TE voltage can be generated by other means of heating. For example, the graphene TE detector is a promising platform for sensing in the challenging THz band.[24,25] Here we focus on the RF response of the graphene TE detector. We couple microwave (MW) radiation capacitively through two bias-tees connected to the graphene device (Fig. 2a inset; more details in Methods). The TE voltage difference $V_d$ between the graphene source and drain leads is read out from the inductive ports of the bias-tees. To reduce standing waves in the circuit, a 10-dB attenuator is inserted after the MW source. We calibrated and measured the input MW power $P_{MW}$ right after the input bias-tee. The measured $V_d$ then allows us to extract the extrinsic MW responsivity of our device $S_{V\_MW} = V_d/P_{MW}$. Figure 2a shows that the device responsivity to 1 GHz MW heating is about 130 ± 10 V/W (more data for measurements up to 50 GHz are shown in the Supporting Information). We note that the observed $S_{V\_MW}$ in Fig. 2a is much smaller than $S_{V\_DC}$ in Fig. 1d under DC heating. This can be understood as a result of impedance mismatch; the graphene load resistance $R_G = 800\ \Omega$ is much larger than 50 Ω and a significant portion of the MW power is reflected rather than delivered to the graphene. The mismatch loss can be estimated by $-10\log_{10}\left[1 - \left(\frac{800-50}{800+50}\right)^2\right] \approx 6.5$ dB. Experimentally we quantify the system loss by performing *S*-parameter measurements using a vector network analyzer (VNA, Keysight N5247A; the measured $S_{11}$, $S_{12}$, $S_{22}$ and $S_{21}$ are shown in Supporting Information). The mismatch loss of the system is given by $-10\log_{10}(1 - |S_{11}|^2 - |S_{21}|^2)$ for port 1, and by $-10\log_{10}(1 - |S_{22}|^2 - |S_{12}|^2)$ for port 2. Figure 2b shows the measured mismatch loss of our sample, which ranges between 8 and 4 dB from low frequency to 50 GHz, consistent with the 6.5 dB estimation from the sample DC resistance. Taking into account the 7.5 dB mismatch loss at 1 GHz, the *intrinsic* responsivity of our detector is then estimated to be 730 ± 60 V/W, which is in reasonable agreement with the results of our DC measurement in Fig. 1d.

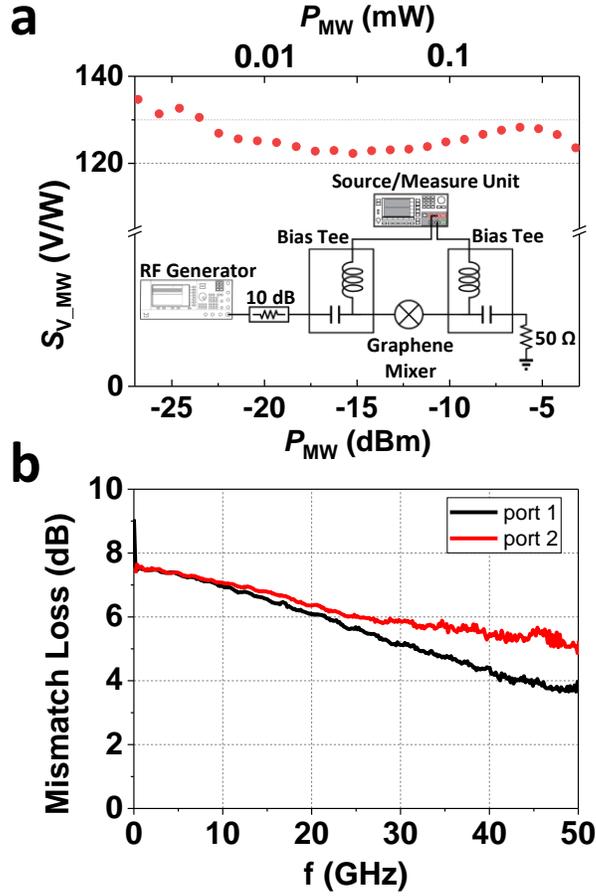

**Figure 2**. (a) Power dependence of direct-detection extrinsic responsivity at 1 GHz. Inset: Microwave direct detection set-up. (b) Mismatch loss of the device calculated from the *S* parameters.

We note that the *extrinsic* responsivity $S_{V\_ext}$ of ~130 V/W or 163 mA/W in Fig. 2a is among the highest observed for graphene high-speed direct-power detectors. In the infrared range, earlier versions of graphene photodetectors for optical communication had $S_{V\_ext}$ of 6.1 mA/W,[18] and later improvements by integration with waveguides reached 35 to 100 mA/W.[6,7,20] In the GHz-THz frequency range, we previously demonstrated a graphene TE detector with a $S_{V\_ext}$ of 4.9 V/W at 1.9 THz.[24] An earlier work using a gated GFET reached 14 V/W at 600 GHz.[26] A more recent GFET design obtained a responsivity of 74 V/W at a frequency of 400 GHz.[15] We anticipate that with future improvements in impedance mismatch, the simple two-terminal graphene device presented here may even become competitive with high-performance diode devices.[27]

The RF detection in Fig. 2 as well as the (sub)THz detections in Refs.15,24,26 yield a DC voltage output (direct power detection) which erases the frequency information of the input signal. We infer from the high response speed of graphene[18,25,28,29] that it should be possible to make a heterodyne RF detector with very large bandwidth, where the RF signal is mixed with a local oscillator (LO), creating radiation at an intermediate frequency (IF) as either an upper side band (USB) or a lower side band (LSB) signal. In this mode of detection, the frequency information of the incident RF signal is retained: $f_{IF}^{USB} = f_{LO} + f_{RF}$ and $f_{IF}^{LSB} = |f_{LO} - f_{RF}|$. RF heterodyne mixers have been demonstrated before with graphene; these are typically three terminal FET devices in which the RF and LO are applied to the gate and the drain separately.[5,12,14] Our design is different in that it is a two-terminal asymmetric device and both RF and LO are applied to the drain electrode. One advantage of this design is the reduction of parasitic capacitance which consequently makes the TE device suitable for achieving much larger bandwidth.

Figure 3a shows our experimental setup for the heterodyne detection. We use a 3 dB RF directional coupler to feed the outputs of two Keysight E8257D analog signal generators into the graphene device through a 10-dB attenuator; one at $f_{LO}$ serving as the local oscillator and the other at $f_{RF}$ as the RF signal. The IF outputs from the graphene device, as well as the transmitted LO and RF, are measured with a calibrated Keysight PXA spectrum analyzer (see Methods for more details).

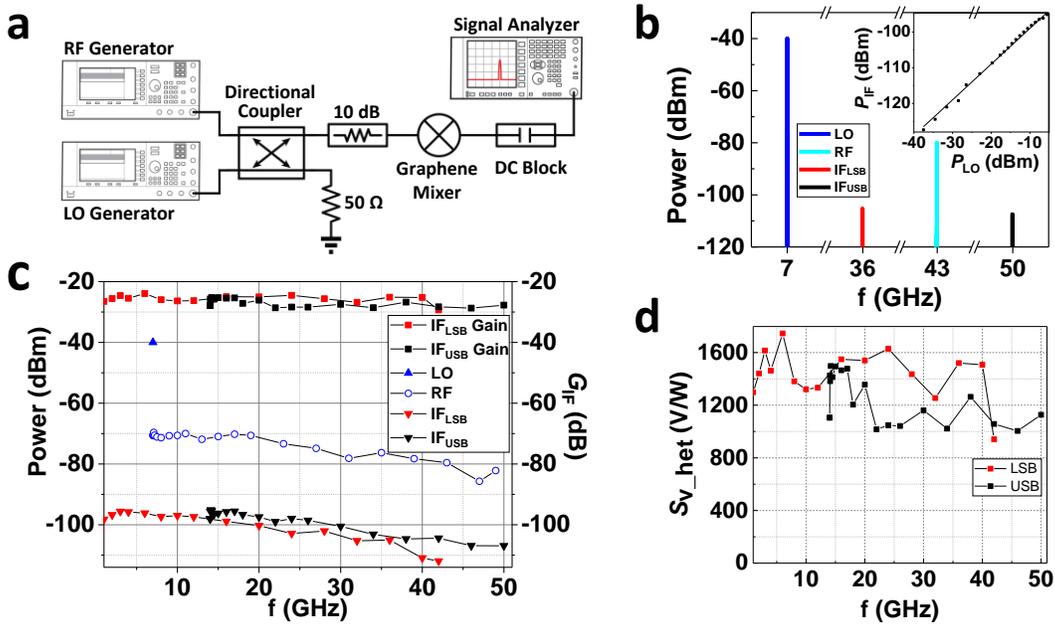

**Figure 3.** (a) RF heterodyne detection setup. (b) Graphene mixer spectrum with LO at 7 GHz and RF at 43 GHz. Inset: IF power as a function of input LO power (the solid line is the linear fit with a fixed slope of 1). (c) Graphene RF mixer bandwidth/speed measurement. The red and black squares are for the conversion gains (right axis). The other symbols are for the transmitted LO, RF, IF$_{LSB}$ and IF$_{USB}$ powers (left axis). (d) Device responsivity assessed from heterodyne mixing.

In Fig. 3b we show a typical heterodyne mixing spectrum. In this measurement, the LO is set at 7 GHz and the RF at 43 GHz; the LSB appears at 36 GHz and the USB at 50 GHz as expected. It is interesting to note that the USB signal has a similar magnitude to the LSB signal, already suggesting that our mixer has very large bandwidth (the slightly smaller height is mostly due to system loss, not the frequency response of the graphene mixer; see below).

Similar to the operation of a hot electron bolometric (HEB) mixer,[30,31] the instantaneous power dissipated in graphene in the presence of LO and RF at different frequencies is given by:

$$P = \frac{(V_{LO}\sin(2\pi f_{LO}t) + V_{RF}\sin(2\pi f_{RF}t))^2}{R_G} \quad (2).$$

Expansion of Eq. 2 results in terms oscillating at the IF frequencies that give rise to TE voltages $V_{IF} = 2\sqrt{P_{LO}P_{RF}}S_V\cos(2\pi(f_{LO} \pm f_{RF})t)$, where $P_{LO} = V_{LO}^2/2R_G$ is the input LO RMS power ($P_{RF}$ is defined similarly). We then find the IF RMS power to be:

$$P_{IF} = \langle\frac{V_{IF}^2}{R_G}\rangle = \frac{4P_{LO}P_{RF}S_V^2\langle\cos^2(2\pi(f_{LO} \pm f_{RF})t)\rangle}{R_G} = \frac{2P_{LO}P_{RF}S_V^2}{R_G} \quad (3),$$

or equivalently the conversion gain is:

$$G_{IF} = \frac{P_{IF}}{P_{RF}} = \frac{2S_V^2 P_{LO}}{R_G} \quad (4).$$

From Eq. 3, $P_{IF}$ is expected to be proportional to $P_{LO}$ at fixed RF power. We have tested this linearity and found that our device is linear over more than 3 decades of change in LO power; see Fig. 3b inset.

We then measured our graphene mixer over as wide a frequency range as allowed by our spectrum analyzer. Fixing the LO at 7 GHz, we tuned the RF from 8 GHz to 50 GHz. Figure 3c shows the measured $IF_{LSB}$ and $IF_{USB}$ powers (down triangle, red and black) together with the LO (up blue triangle) and RF (open blue circle) power. We note that $IF_{LSB}$ and $IF_{USB}$ powers are about -95 dBm at low frequencies and decrease to about -110 dBm at high frequencies; a similar ~15 dB decrease of power is also observed for the RF. The similar IF and RF power decrease indicates that this roll-off is due to measurement system loss, not the frequency response of the graphene mixer. Indeed, using the definition in Eq.

4, we find the conversion gain of our graphene device to be $G_{IF}$ = -27 ± 3 dB over the whole 50 GHz range (Fig. 3c, red and black squares).

This bandwidth is much larger than previous graphene RF mixers[5,12,14,16] and is highly desirable for wideband applications. The wide bandwidth ($B$) indicates that the thermal time constant of the heated electrons in graphene is shorter than $\tau = 1/(2\pi B) = 3$ ps. Similar high speed / wide bandwidths have been demonstrated in graphene mixer experiments with photons in the optical communication bands.[19,20] We note that the wide IF bandwidth in all these cases is consistent with predictions that the intrinsic device speed for graphene detectors that rely on electron heating is at least 260 GHz.[32]

We can also compare the bandwidth of our graphene mixer with that of HEB mixers. HEB mixers with NbN superconducting devices have been limited to about 4 GHz bandwidth[33] but recent results on $MgB_2$ superconducting HEB mixers have increased this to about 10 GHz.[34] Two-dimensional electron gas (2DEG) HEB mixers, also operating at cryogenic temperatures, have demonstrated 40 GHz bandwidth by exploiting ballistic transport.[35]

We can make use of the data in Fig. 3c and the mismatch loss measurement in Fig. 2b to extract the voltage responsivity $S_{V\_het}$ using Eq. 4. As shown in Fig. 3d, we find $S_{V\_het}$ = 1300 ± 300 V/W over 50 GHz with experimental uncertainty less than 3 dB, agreeing well with the DC measurements in Fig. 1d. This consistency validates our understanding of graphene heterodyne detection mechanism as described in Eqs. 2-4, and further confirms that the DC rectification provides a useful gauge to estimate device performance.

The fourth technique we use to measure the TE responsivity of our device is through 2nd harmonic generation. Graphene is a material with an inversion center. From standard optical selection rules, 2nd harmonic generation in graphene is forbidden while the 3rd harmonic is allowed.[22] Our device, however, displays the opposite behavior: in Fig. 4a, we excited our sample at 16 GHz, and observed that the 2nd harmonic signal at 32 GHz is more than two orders of magnitude stronger than the 3rd harmonic signal at 48 GHz. We further performed power dependence in Fig. 4b and found that the 2nd harmonic signal indeed has quadratic input power dependence.

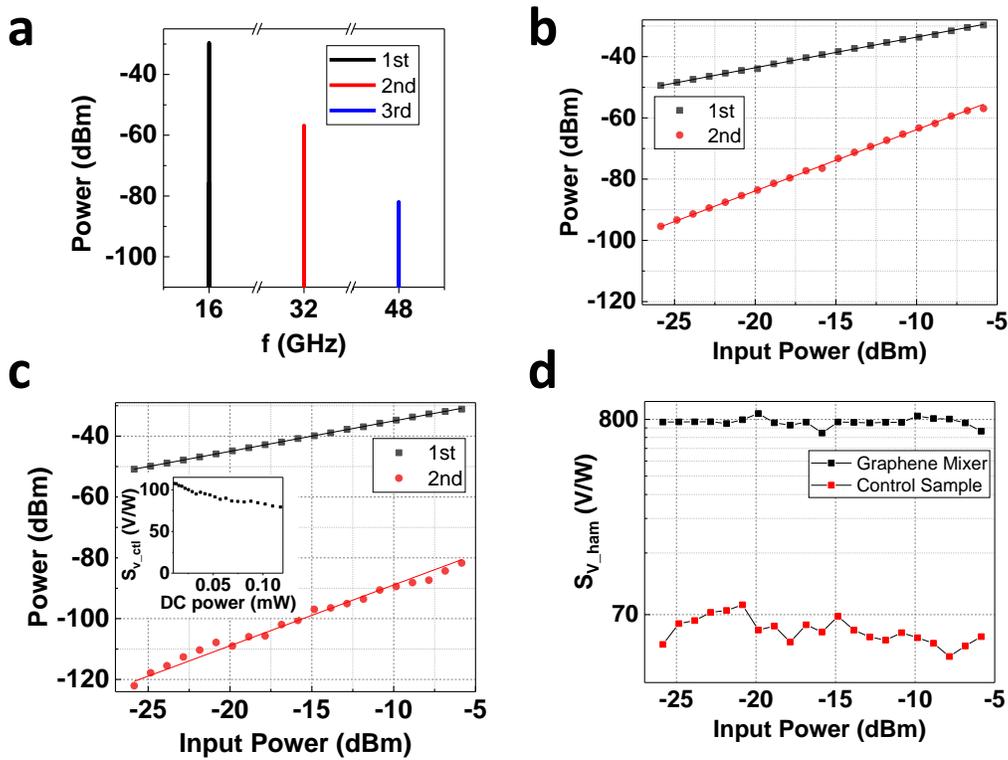

**Figure 4.** (a) The spectrum for harmonic signals generated in the graphene mixer device. The fundamental is set at 16 GHz. (b) Input power dependence of the 1st and 2nd harmonic signals; the solid lines are linear fits with fixed slopes of 1 and 2. (c) Same measurements as in (b) performed on the control sample. Inset: DC responsivity of the control sample. (d) Calculated responsivity from the second-harmonic conversion gain as a function of input power.

While forbidden for standalone graphene, the appearance of a 2nd harmonic signal is not so strange in light of the capability of our device to mix the RF and LO inputs. The 2nd harmonic signal we observe can be viewed as the self-mixing of RF electric fields in the device due to the asymmetric contacts. To understand this phenomenon quantitatively, we revised the formulation in Eqs. 2-4. Here, the graphene is driven by MW at a single frequency and the instantaneous power is given by: $P = V^2 \sin^2(2\pi ft)/R_G = V^2(1-\cos(4\pi ft))/2R_G$. The second term is responsible for the 2nd harmonic TE voltage: $V_{2f} = P_f S_V \cos(4\pi ft)$. We find

$$P_{2f} = \frac{P_f^2 S_V^2}{2R_G} \qquad (5).$$

Similar to the application of Eq. 4 to find $S_{V\_het}$, here we make use of Eq. 5 to extract $S_{V\_harm}$. This is shown in Fig. 4d (black squares). The extracted value of $S_{V\_harm} = 800 \pm 200$ V/W is in reasonable agreement with the values obtained from DC heating, RF heating, and RF mixing in Figures 1-3.

To further validate our understanding of this unusually strong 2nd harmonic signal, we have fabricated a control device with similar DC resistance (750 Ω) but a smaller TE responsivity of about $90 \pm 20$ V/W (Fig. 4c inset). Interestingly its 2nd harmonic signal is much smaller, about two orders of magnitude weaker. This can be understood from Eq. 5, where it is shown that the 2nd harmonic intensity is expected to scale as $S_V^2$. The two orders of magnitude weaker 2nd harmonic generation is in excellent agreement with the control sample's ~10 times smaller $S_V$. Lastly, we plotted the estimated $S_{V\_harm}$ for the control device in Fig. 4d (red squares), which gives $70 \pm 10$ V/W, matching well with its DC responsivity in Fig. 4c inset.

The graphene RF device we developed here represents an important advance catching up with corresponding developments in the infrared optical frequencies.[6,7,18–20] The device is operating efficiently as a direct detector, exhibiting an external responsivity as high as 130 V/W (163 mA/W) over a large dynamic range, which is among the best for superfast graphene detectors. Heterodyne measurements demonstrated an intrinsic conversion loss of 27 dB and a bandwidth of more than 50 GHz, which are highly promising. It is of interest to compare our graphene mixer with existing technologies based on CMOS, SiGe and III-Vs (GaAs and InP). Currently CMOS technology is the most feasible for large scale systems and has the lowest cost. As was shown by Khamaisi et al.,[36] for 65 nm CMOS the conversion loss is in the range of 23-25.5 dB for LO frequencies from 220-300 GHz, comparable to our graphene mixer for intrinsic conversion loss. Bandwidths are 20-30 GHz, smaller than graphene. At shorter wavelengths beyond the transistor cutoff frequency $f_{max}$, the conversion loss increases steeply. By incorporating LNAs (low-noise amplifiers) for RF and/or IF, more expensive integrated RF receivers have achieved better performances: SiGe, 15 dB conversion gain and 28 GHz bandwidth (including a preamplifier) at 220 GHz;[37] GaAs metamorphic high electron-mobility transistors (mHEMTs), 3.5 dB gain and more than 10 GHz bandwidth at 220 GHz;[38] InP, the best performer, reaching 26 dB gain at 298 GHz with a 3-dB bandwidth of 20 GHz.[39] Our graphene RF mixer has larger bandwidth than the above technologies and has ample room for improvement in terms of operation frequency and conversion gain. The TE detection relies on absorption of RF power in a device with very low parasitic reactance. This lends its operation to translation to much higher frequencies, such as the sub-THz semiconductor devices as discussed above,[36–39] as well as further up to several THz, related to and improving upon the device we developed before.[24] The conversion gain can be improved by applying a gate voltage,[24] lowering the device impedance and optimizing the device asymmetry. With its successful integration with CMOS,[7] we envision that graphene RF mixers can be integrated with high performance RF and IF LNAs[37–39] to further improve their performance, promising for emerging applications in high-speed communication systems at a few hundred GHz,[40] as well as in thermal imaging systems in the THz range.[41,42]

## Methods:

### Sample fabrication.

The detector devices were fabricated on CPW lines on a high resistivity silicon substrate with pads designed for microwave probing, see Fig.1 a-c. The CPW was fabricated following Ref. 43 in two steps: (1) pad frame and feed lines, (2) contact pads. The pad frame and coplanar feed-lines were defined by photolithography with 5 nm thick Ti and 80 nm thick Au deposited on the sample successively in an electron beam evaporator. The contact pads were then patterned in a similar way using photolithography with 1 μm thick Cu and 500 nm thick Al deposited via magnetron sputtering. The fabricated structure consists of two GSG ports between which a 50 Ω CPW structure was formed. The graphene detector was placed in series with the Au center conductor of the CPW. The detector was fabricated with asymmetric contacts, similar (but not identical) to the device described in our work on a THz detector.[24] The graphene and hBN samples we used in this study are all made by mechanical exfoliation. The graphene flake was confirmed to be monolayer by Raman spectroscopy.[44] We first made a BN/graphene/BN sandwich sample by the dry transfer method[45] and then transferred the sandwich to the gap in between the central leads of the CPW, with graphene covering one side of the Au electrode. The other side of the sandwich is patterned with a two-layer PMMA mask, dry-etched using $CHF_3/O_2$ plasma and subsequently contacted to the left electrode with a graphene edge contact using palladium sputtering, similar to Ref. 23. This introduces asymmetry in the device that enables microwave detection and second-harmonic generation. The control sample is made by hBN covered graphene instead of a sandwich. To have similar resistance for easy comparison, the control sample's channel is about three times wider.

**RF direct detector measurements.**

We couple MW signals into the graphene device via bias-tees connected to GSG probes with 100 μm pitch. The device is fed from a Keysight E8257D signal source, while the detected voltage is picked up on the center conductor and measured through two bias-tees (Anritsu K251) by a Keysight B2901A Precision Source/Measure Unit used as a voltmeter (see Fig.2a). The available MW power after the input bias-tee was measured with a Keysight N1913A power meter using a Keysight N8488A power sensor and was typically from -28 dBm to -2 dBm. The signal generator output has an upper frequency of 50 GHz. By normalizing the thermoelectric voltage to the available input microwave power before the graphene device, we find the MW responsivity in Fig. 2a.

**RF heterodyne- and self-mixing measurements.**

For heterodyne measurements, the device is fed with LO and RF power at different GHz frequencies with two Keysight E8257D analog signal generators through a 3-dB RF directional coupler and a 10-dB attenuator into the probe station. The LO power is set at -8 dBm at the input of the device. The RF power is set 30 dB lower at -38 dBm. The LO, RF and IF powers are measured from the output side of the device with a calibrated Keysight PXA spectrum analyzer. We verified that mixing of the LO and RF internal to the spectrum analyzer produces negligible IF power. For self-mixing measurements, we simply apply only one microwave signal to the detector through a 10-dB attenuator and measure the transmitted power of the harmonic signal with the spectrum analyzer. The input power is swept from -6 dBm to -26 dBm at the device input at 16 GHz for the two devices that were measured.

**Supporting Information**

*S*-parameter measurements of the device, the frequency dependence of the RF direct detector responsivity, and additional data on higher harmonic component measurements are found in the Supporting Information.

**Author information**

Corresponding Author:

[†]Jun Yan    E-mail: yan@physics.umass.edu

ORCID
J. Yan: 0000-0003-3861-4633

**Notes**

The authors declare no competing financial interest.

**Acknowledgements**

This project was supported by the National Science Foundation under Grant Number ECCS-1509599. M. C. C. is supported by the Raytheon Integrated Defense Systems Engineering Advanced Studies Program Fellowship.

**Table of Contents Graphic**

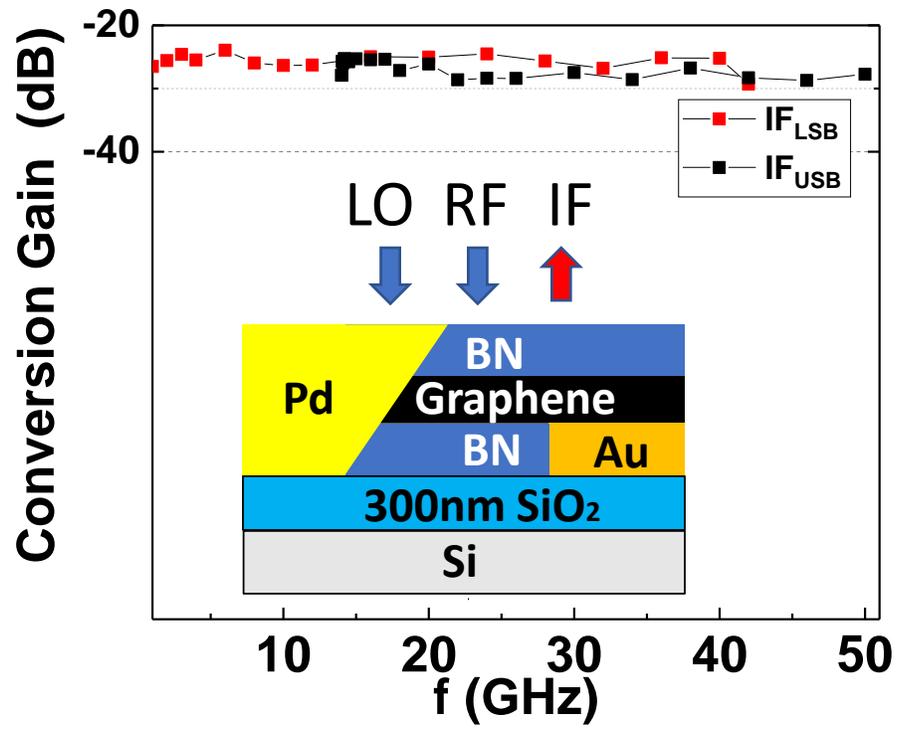